\documentstyle[12pt]{article}
\def\ra{\rightarrow}

\def\be{\begin{equation}}
\def\ee{\end{equation}}
\def\bea{\begin{eqnarray}}
\def\eea{\end{eqnarray}}

\begin{document}

\begin{titlepage}
\thispagestyle{empty}
\begin{center}
{ \bf
{EXPERIMENTAL REALIZATIONS OF INTEGRABLE REACTION-DIFFUSION 
PROCESSES IN BIOLOGICAL AND CHEMICAL SYSTEMS}
\footnote{Invited lecture given at the $7^{th}$ Nankai Workshop on 
Symmetry, Statistical Mechanics Models, and Applications 
at Nankai University, Tianjin (August 1995). To appear in the
Proceedings, eds. F.Y. Wu and M.L. Ge, (World Scientific, Singapore, 1996)}
}\\[25mm]

{ {\sc
Gunter M. Sch\"utz}
}\\[8mm]

\begin{minipage}[t]{13cm}
\begin{center}
{\small\sl
Department of Physics, University of Oxford\\
Theoretical Physics, 1 Keble Road, Oxford OX1 3NP, UK
}
\end{center}
\end{minipage}
\vspace{40mm}
\end{center}
{\small
Stochastic reaction-diffusion processes may be presented in terms of integrable quantum chains and can be used to describe various biological and chemical 
systems. Exploiting the integrability of the models one finds in some cases 
good agreement between experimental and exact theoretical data. This is shown 
for the Rubinstein-Duke model for gel-electrophoresis of DNA, the asymmetric 
exclusion process as a model for the kinetics of biopolymerization and
the coagulation-diffusion model for exciton dynamics on TMMC chains.
}
\\
\vspace{5mm}\\
\end{titlepage}

\section{Introduction}

It has been realized in recent years that the stochastic time evolution
of many one-dimensional reaction-diffusion processes can be
mapped to integrable quantum chains. 
This insight has made available the
tool box of integrable models for these interacting particle systems
far from equilibrium and has led to many new exact results for their dynamical 
and stationary properties. It is also amusing to note that the
Hamiltonians for such systems are, in general, not hermitian and therefore
from a quantum mechanical point of view not interesting. The interpretation
as time evolution operators for stochastic dynamics thus extends the physical 
relevance of integrable systems to non-hermitian models.

The relationship
between stochastic dynamics and quantum chains is conceptually very simple:
At any given time $t$ the state of the system is completely described by a
probability distribution $f(\underline{n};t)$ for the stochastic variables
$\underline{n}$. The time evolution of this distribution is governed by
a master equation which expresses the probability of finding the system at 
time $t+dt$ in a given configuration in terms of the probability distribution 
at time $t$ through a first order differential equation in the time variable. 
Such a master equation can be expressed in a ``quantum
Hamiltonian formalism'' by mapping each state of the system to a basis 
vector in a suitable vector space $X$. In this mapping the probability 
distribution at time $t$ becomes a vector $|\,f(t)\,\rangle$ and the master 
equation takes the form
\be
\frac{d}{d t} |\,f(t)\,\rangle = - H |\,f(t)\,\rangle
\label{1}
\ee
where $H$ is a suitably chosen linear map acting on $X$.
For many systems of interest, $H$ is the Hamiltonian of some integrable
quantum chain.\footnote{See e.g. the paper on
Stochastic Reaction-Diffusion Processes, Operator Algebras and Integrable
Quantum Spin Chains in this volume and references therein.}

A severe constraint to the application of such models to experiments seems to 
be the fact that they are all one-dimensional. However, it turns out that 
for many situations e.g. involving polymers or traffic models, a 
one-dimensional description of the stochastic dynamics is appropriate.
Furthermore there are systems where only the projection of the system 
onto one space coordinate is of actual experimental interest. In these
cases integrable reaction-diffusion processes can and do play a role. 

Here we will briefly review three different systems where integrable models
are of experimental relevance: Firstly the Rubinstein-Duke model for the
description of gel-electrophoresis of DNA (Sec.~2), then the asymmetric
exclusion process with open boundary conditions for the kinetics of 
biopolymerization on nucleic acid templates (Sec.~3) and the one-species
coagulation-diffusion model for exciton dynamics on laser excited TMMC chains 
(Sec.~4).

\section{Gel-Electrophoresis of DNA}

A widely used and simple method for the separation of DNA fragments of 
different length is DC gel-electrophoresis. The DNA mixture to be separated
is introduced into a gel matrix. Since the DNA is charged, it will move
in a constant electric field $E$ with velocity $v(E,N)$ where $N$ is the
length of the fragment. After some time fragments of different length will have 
travelled a distance in the gel depending on their length and can therefore be 
separated. Clearly it would be desirable to have a quantitative 
understanding of the motion of DNA in gels.

Based on the earlier concepts of the confining tube\cite{Edwards} and 
of reptation\cite{de Gennes} Rubinstein\cite{Rubin} and Duke\cite{Duke}
introduced a simple model for the motion of a polymer in a gel matrix.
In this model the gel is idealized
by a regular cubic lattice where the cells are the pores of the gel
through which the polymer reptates. The polymer itself is represented by
a string of reptons, the number of which is the length of the polymer
divided by its persistence length. These reptons hop stochastically from
pore to pore according to rules based on the mechanism
of reptation and assuming local detailed balance. 
Since in electrophoresis only the
average velocity of the center of mass in field direction is of interest,
one can project the motion of the reptons onto this direction. Some
mappings that we will not describe here lead finally to a lattice gas
model representing the relative motion of all reptons in field direction.
The motion perpendicular to the field is diffusive with a diffusion 
constant\cite{VWD,Prae} $D=1/3N^2$ entering the drift velocity $v(E,N)$ for 
small $E$ through the Nernst-Einstein relation\cite{VWD,L} $v=DNE$.

Here we will not give the details of the mapping to the lattice gas model but
just state the resulting lattice gas dynamics:\cite{BMW} 
There are two kinds of particles, $A$ and $B$, moving on a lattice of
$L=N-1$ sites and each site can occupied by at most one particle, $A$ or
$B$. $A$-particles hop to right (left) with rate $q$ ($q^{-1}$) if the site is 
unoccupied. Here $\ln(q)$ is the energy gain when a repton moves into a pore 
in field direction. On site 1 of the chain $A$-particles are created 
(annihilated) with rate $q$ ($q^{-1}$), while on site $L$ they are
annihilated (created) with rate $q$ ($q^{-1}$). For the $B$-particles
the same rules hold, but with $q$ and $q^{-1}$ interchanged.
It is easy to show that the average drift velocity $v(E,N)$ is the difference
between the stationary current $j_A(E,N)-j_B(E,N)$ of $A$ particles and  
$B$ particles.

The stationary distribution of the system is not known except in the
periodic system\cite{LK} which does not have an interpretation in terms of 
polymers moving through a gel. However, extensive Monte-Carlo studies\cite{BMW} 
have provided a good and reliable knowledge of $v$ in the framework of the
model. The surprise is that these results are in 
excellent agreement with experimental data\cite{ex1,BS}. This gives confidence 
that despite all its simplifications, the Rubinstein-Duke model captures the 
essential physical processes involved and allows for reliable predictions in
real gel-electrophoresis. 

In order to make contact with integrable models we write the master equation
of the process in the quantum Hamiltonian formalism.
The stochastic time evolution of the system is given by the 
Hamiltonian of a three-states quantum chain\cite{BS}
\be
H(\alpha,q) = b_1(\alpha,q) + b_L(\alpha,q^{-1}) + \sum_{i=1}^{L-1} u_i(q)
\ee
where $b_i(\alpha,q) = \alpha q (1 - n_i^A - a_i^{+} - b_i)  + \alpha q^{-1} 
(1 - n_i^B - a_i - b_i^{+})$ and 
$u_i(q) =  q (n_i^A n_{i+1}^0 + n_i^0 n_{i+1}^B - a_i a_{i+1}^{+} - b_i^{+} 
b_{i+1} ) + q^{-1} (n_i^0 n_{i+1}^A + n_i^B n_{i+1}^0 - a_i^{+} a_{i+1} 
 -  b_i b_{i+1}^{+} )$. 
Here $n_i^A \equiv E_i^{11}$, $n_i^B \equiv E_i^{33}$ and
$n_i^0 = 1-n_i^A-n_i^B \equiv E_i^{22}$ are projection operators on states
with an $A$-particle, vacancy and $B$-particle resp. on site $i$. The operators
$a_i \equiv E_i^{21}$, $a_i^{+} \equiv E_i^{12}$, $b_i \equiv E_i^{23}$,
and $b_i^{+} \equiv E_i^{32}$ are annihilation and 
creation operators for $A$- and $B$ particles. $E_i^{jk}$ is the $3\times 3$
matrix with matrix elements $(E_i^{jk})_{\alpha,\beta}=\delta_{j,\alpha}
\delta_{k,\beta}$ acting on site $i$. The factor $\alpha$ takes into account
the possibility of a different mobility of the end-reptons compared to
those in the bulk.

Nothing is known about the integrability of the model in non-zero field.
However, if no field is applied ($q=1$) and if the ends of the polymer are 
fixed in the gel ($\alpha=0$, e.g. by making a chemical bond with an immobile 
particle), then $H$ is integrable. In this case the model 
describes the internal random fluctuations of the polymer within the gel. That 
$H(0,1)$ is integrable can be seen by verifying that for $\alpha=E=0$ the 
Hamiltonian for both the isotropic spin-1/2 Heisenberg chain with open boundary 
conditions and the Rubinstein-Duke model have the form $H=\sum_{i=1}^{L-1} u_i$
where the $u_i$ satisfy the same Temperley-Lieb algebra $u_i^2 = 2 u_i$,
$u_i u_{i\pm 1} u_i = u_i$, $[u_i,u_j]=0$ for $|i-j|\geq 2$. 
Using the Bethe ansatz one can compute the relaxation of
the DNA to equilibrium (where each configuration is equally probable). 

If the ends of the polymer are not kept fixed, then the model has at least
an integrable subspace with a spectrum which is identical to that of the
isotropic Heisenberg chain with non-diagonal, symmetry breaking boundary
fields. This can be shown by using a similarity transformation on
$H$ and projecting on one of its invariant subspaces\cite{SY}. In this case
one can use the integrability to obtain the relaxation of the distribution
of vacancies. This gives the density of reptons per pore in a freely diffusing
polymer. 

Remarkably the model predicts that there is no band collapse
if one pulls only at one end of the polymer rather than at each repton
with some external field\cite{BS}. 
The velocity is then always length dependent and asymptotically
given by the exact expressions $v = E/{3N^2}$ for $E \rightarrow 0$ and
$v = 1/(3N-5)$ for $ E \rightarrow \infty$.
In this case $H$ consists of the integrable zero-field bulk part and boundary
terms $b_1(1,q) + b_L(1,1)$. Whether this model is integrable is not clear and 
it would be interesting to ask generally which integrable boundary conditions 
one can obtain for the model with vanishing bulk field\cite{Skl}.

\section{Kinetics of Biopolymerization}

Back in 1968 MacDonald et al.\cite{bio1,bio2,bio3} studied the kinetics of 
biopolymerization on nucleic acid templates. The mechanism they try to describe 
is (in a very simplified manner) the following: Ribosomes attach to the 
beginning of a messenger-RNA chain and ``read'' the genetic information which 
is encoded in triplets of base pairs by moving along the m-RNA.\footnote{The 
m-RNA is a long molecule made up of such consecutive triplets.} 
At the same time the 
ribosome adds monomers to a biopolymer attached to  it: Each time a unit of 
information is being read a monomer is added to a biopolymer attached to the 
ribosom and which is in this way synthesized by the ribosom. 
After having added the 
monomer the ribosom moves one triplet further and reads again. So in each 
reading step the biopolymer grows in length by one monomer. Which monomer is 
added depends on the genetic information read by the ribosom. The ribosoms are 
much bigger than the triplets on the m-RNA, they cover 20-30 of such triplets. 
Therefore neighbouring ribosomes sitting at the same time on the m-RNA cannot 
simultaneously read the same information. Furthermore they cannot overtake 
eachother: If a ribosom sits at a particular 
place on the m-RNA and does not (temporarily) proceed further (e.g. because no 
appropriate monomer has been found in the surrounding medium for the 
polymerization process), then an oncoming 
ribosom from behind will stop until the first has eventually moved on.

In order to describe the kinetics of this process MacDonald et al. 
introduced the following
simple model. The m-RNA is represented by one-dimensional lattice of $L$ sites 
where each lattice site represents one triplet of base pairs. The ribosom is
a particle covering $r$ neighbouring sites (for real systems  $r
=20\dots 30$) but moving by only one lattice site in each (infinitesimal) time 
step with a constant rate $p$. These particles interact via hard-core repulsion, 
i.e. there is no long range interaction, but there is also no overlap of 
ribosomes. In principle one can also allow for back-hopping with a non-vanishing
rate $q$. At the beginning of the chain particles are added with rate $\alpha p$
and at the end of the chain they are removed with rate $\beta p$. Again, one
may also allow for a removal rate $\alpha' q$ at the beginning of the chain
and an addition rate $\beta' p$ at the end of the chain.

In the idealized case $r=1$ this model became later known as the
asymmetric exclusion process with open boundary conditions\cite{ZS}. Its steady
state was first studied using a mean-field approach\cite{bio2}. Then in a 
following paper\cite{bio3} the generalized case $r > 1$ was studied 
numerically and compared to experimental data on the stationary density
distribution of ribosomes along the chain. These were found to be consistent 
with the results obtained from the model with $q=0$ and $\alpha=\beta<p/2$.
Furthermore it turned out that the phase diagram for general $r$
is similar to the much simpler case $r=1$ in the sense that there
are three distinct phases, a low density phase, a high density phase
and a maximal current phase (see below). These observations encourage
to use the asymmetric exclusion process as a simple but in certain aspects 
realistic model for this biological system.

The experimentally relevant case is the
phase transition line from the low-density phase to the high density phase.
On this phase transition line the mean-field and numerical calculations
predict a region of low density of ribosomes from the beginning of the
chain up to some point where the density suddenly jumps (over a few lattice
sites) to a high density value.\footnote{This description of the stationary 
mean-field density profile describes correctly the situation for $r=1$, 
but disregards a more complicated sublattice structure for $r > 1$.
However the figures provided by MacDonald et al.\protect\cite{bio3} suggest
that the description remains qualitatively correct when one averages
over this sublattice structure.} In the maximal current phase 
$\alpha,\beta>p/2$ mean-field predicts a power law decay of the stationary 
density to a constant bulk value, $\rho(x)-\rho_{bulk} \propto x^{-1}$, where
$x$ measures the distance from the ends of the chain. In this case the
polymerization determines the profile rather the initialization and release
rates $\alpha,\beta$.

These predictions and the apparent experimental relevance of the model make an 
exact solution of at least the simple case $r=1$ desirable. The 
stochastic dynamics of the model are given in the quantum Hamiltonian formalism
by the integrable Hamiltonian of the anisotropic spin-1/2 Heisenberg chain with 
non-diagonal boundary fields\footnote{The 
equivalence to the Heisenberg chain is more obvious after the similarity 
transformation $\Phi$ given in the first lecture elsewhere in this 
volume.}\cite{bound}
\bea
H & = & - \alpha p \left[ s^-_1 - (1-n_1) \right] 
        - \beta p \left[ s^+_1 - n_1 \right] \nonumber \\
 & & - \sum_{i=1}^L \left[ p \left( s^+_i s^-_{i+1} - n_i(1-n_{i+1}) \right) +
                    q \left( s^-_i s^+_{i+1} - (1-n_i)n_{i+1} \right) \right].
\eea
Here $s^{\pm}=(\sigma^x\pm i \sigma^y)/2$ and $n=(1-\sigma^z)/2$.
For $\alpha=\beta=0$ this reduces to the $SU(2)_q$ symmetric quantum chain
with diagonal boundary fields which can be solved by the coordinate or
algebraic Bethe ansatz. However, the boundary fields given here break the
$U(1)$ symmetry of the model and other approaches are necessary to find
at least the steady state of the system, i.e. the ground state of $H$ with
(by construction) energy 0. In what follows we will consider only $q=0$.
We set $p=1$ which is no loss in generality since it sets only the time scale
of the process. 

The breakthrough to the exact solution came only more than 20 years after
the work on biopolymerization and independently of it\cite{DDM,SD,DEHP}. 
It turned out that the solution of the master equation of a system of $L$ sites 
can be recursively expressed in terms of the solution for $L-1$ sites\cite{DDM}.
The exact solution obtained from the solution of these recursion relations 
reproduces the three phases predicted by mean field, but
shows more structure inside the low- and high-density phases\cite{SD}. 
This reveals
an intricate interplay between two correlation length which determine the
phase diagram and the nature of the phase transitions. In particular,
it turns out that the correlation length on the phase transition line
between the low-density phase and the high-density phase is infinite, which
is incompatible with the mean field result. The exact solution gives a linearly
increasing density profile rather than the sharp shock predicted by mean 
field\cite{SD}. This can be explained by assuming that a sharp shock exists, 
but, due to current fluctuations, performs
a random walk along the lattice. Therefore, if one waits long enough,
the shock will have been at each lattice with equal probability. This picture
yields a linearly increasing density and is confirmed by an exact
solution of dynamical properties of a related exclusion process with
deterministic bulk dynamics\cite{sch}. 
What one therefore expects for an experimental
sample is indeed a region of low density of ribosoms followed by a sharp
transition to a region of high density of ribosoms as found experimentally. 
This rapid increase can be
anywhere on the m-RNA, but with a probability distribution given by
the effective
initialization and release rates $\alpha,\beta$. If $\alpha=\beta$ the
distribution of shock position would be constant over the lattice, otherwise
exponential on a length scale\cite{SD}
$\xi=1/(\ln{[\alpha(1-\alpha)/\beta(1-\beta)]})$.
If $\alpha,\beta > 1/2$, i.e. when polymerization determines the dynamics, 
then the exact solution predicts an algebraic decay of the
density to its bulk value 1/2 with exponent $b=1/2$ rather than $b=1$ predicted
by mean field.

\section{Exciton Dynamics on Polymer Chains}

Finally I would like to discuss briefly an experiment in which excitons
on polymer chains are created by laser excitations and then hop on the
chain and coagulate when they meet.
The carrier substance is $(CH_3)_4 N MnCl_3$ (TMMC). The particles are excitons
of the $Mn^{2+}$ ion and move along the widely separated $MnCl_3$ chains.
A single exciton has a decay time of about
$0.7 ms$. The on-chain hopping rate is $10^{11}-10^{12} s^{-1}$.
If two excitons arrive on the same $Mn^{2+}$ ion, they undergo a 
coagulation reaction $A+A\ra A$ with a reaction time 
$\approx 100fs$ \cite{Kroon93}. 

It has been suggested to describe this process by the coagulation-diffusion 
model on a one-dimensional lattice\cite{Simon95,hos95} which through a
similarity transformation is equivalent to the diffusion limited pair 
annihilation process\cite{Krebs95}. In this model each lattice site may be 
occupied by at most one particle. These particles hop with rates $D$ to the
right or left nearest neighbouring site resp. if this site is vacant and both 
annihilate with rate $\lambda$ if it is occupied. The annihilation rate
is equal to the coagulation rate of the original process. Since the experimental
data suggest that the coagulation is approximately instantaneous, one finds
$\lambda=D$\cite{hos95}. In the quantum Hamiltonian formalism the stochastic 
time evolution of this transformed process is then given by the Hamiltonian
\be
H = - D \sum_{k=1}^L\left(s^+_k s^-_{k+1} + s^-_k s^+_{k+1} 
                         + s^+_k s^+_{k+1} - \sigma^z_k + 1  \right). 
\ee
Here $D$ sets the time scale for the diffusion. The finite life time $\tau$
of the excitons is much larger than $D^{-1}$, thus a decay term 
$\tau^{-1} \sum (s^+_i-n_i)$ is neglected. This Hamiltonian can be turned by a 
Jordan-Wigner transformation into an integrable free fermion system. One then
finds\cite{racz} that the average density of excitons decays algebraically in time 
with an exponent $x=1/2$ in good agreement with the experimental result 
$x=0.48(2)$\cite{Kroon93}. 
The model predicts also an independence of the amplitude of the decay for long 
times from the initial density as seen in the experiment\cite{Kroon93,hos95}.

\section*{Acknowledgments}
It is a pleasure to acknowledge fruitful collaborations with G.T. Barkema,
E. Domany, M. Henkel and E. Orlandini and to thank the organizers of this 
workshop
for their kind invitation. This work was supported by an EEC fellowship
under the Human Capital and Mobility program.


\begin{thebibliography}{99}

\bibitem{Edwards}
S.F. Edwards, Proc. Phys. Soc., London {\bf 92}, 9 (1967).

\bibitem{de Gennes}
P.G. de Gennes, J. Chem. Phys. {\bf 55}, 572 (1971).

\bibitem{Rubin}
M.~Rubinstein, Phys. Rev. Lett. {\bf 59}, 1946 (1987).

\bibitem{Duke}
T.A.J.~Duke, Phys. Rev. Lett. {\bf 62}, 2877 (1989).

\bibitem{VWD}
B.Widom, J.L.Viovy and A.D. D\'efontaines,
J.Phys. I France {\bf 1}, 1759 (1991).

\bibitem{Prae} M. Pr\"ahofer, Thesis University Munich (1994).

\bibitem{L}
J.M.J.~van Leeuwen, J.~Phys.~I~France {\bf 1}, 1675 (1991).

\bibitem{LK}
J.M.J.~van~Leeuwen and A.~Kooiman, Physica A {\bf 184}, 79 (1992).

\bibitem{BMW}
G.T. Barkema, J. Marko and B. Widom, 
Phys. Rev. E {\bf 49}, 5303 (1994).

\bibitem{ex1}
G.T. Barkema, C. Caron, and J.F. Marko, to appear in {\it Biopolymers}.

\bibitem{BS}
G.T. Barkema and G.M. Sch\"utz, Oxford preprint.

\bibitem{SY} G.M. Sch\"utz and C.M. Yung, in preparation.

\bibitem{Skl}
E. K. Sklyanin, J. Phys. A, \underline{21}, 2375 (1988).

\bibitem{bound}
H.-J. de Vega and A. Gonzalez-Ruiz, J. Phys. A \underline{27}, 6129 (1994).

\bibitem{bio1}
A.C. Pipkin and  J. H. Gibbs, Biopolymers {\bf 4}, 3 (1966).

\bibitem{bio2}
J.T. MacDonald, J. H. Gibbs and A. C. Pipkin, Biopolymers {\bf 6}, 1 (1968).

\bibitem{bio3}
J.T. MacDonald and J. H. Gibbs, Biopolymers {\bf 7}, 707 (1969).

\bibitem{ZS}
R.K.P. Zia and B. Schmittmann, in {\em Phase Transitions and Critical Phenomena},
eds. C Domb and J. Lebowitz (Academic Press, London, 1995).

\bibitem{DDM} B. Derrida, E. Domany and D. Mukamel, J. Stat. Phys. {\bf 69}, 
667 (1992).

\bibitem{SD} G. Sch\"utz and E. Domany, J. Stat. Phys. {\bf 72}, 277 (1993).

\bibitem{DEHP}
B. Derrida, M.R. Evans, V. Hakim and V. Pasquier,
J. Phys. A {\bf 26}, 1493 (1993).

\bibitem{sch} G. Sch\"utz, Phys. Rev. E {\bf 47}, 4265 (1993).

\bibitem{Kroon93} R. Kroon, H. Fleurent and R. Sprik, Phys. Rev. {\bf E47},
2462 (1993).

\bibitem{Simon95} H. Simon, J. Phys. A. {\bf 28}, 6585 (1995).

\bibitem{hos95} M. Henkel, E. Orlandini and G.M. Sch\"utz,
J. Phys. A {\bf 28}, 6335 (1995).

\bibitem{Krebs95} K. Krebs, M. P. Pfannm\"uller, B. Wehefritz and
H. Hinrichsen, J. Stat. Phys. {\bf 78}, 1429 (1995)

\bibitem{racz}
Z. Racz, Phys. Rev. Lett. {\bf 55}, 1707 (1985).



\end{thebibliography}
\end{document}